\documentclass[12pt]{article}

\newcommand{\be}{\begin{equation}}
\newcommand{\ee}{\end{equation}}
\newcommand{\bel}[1]{\begin{equation}\label{#1}}
\newcommand{\bea}{\begin{eqnarray}}
\newcommand{\eea}{\end{eqnarray}}
\newcommand{\ba}{\begin{array}}
\newcommand{\ea}{\end{array}}

\newcommand{\nn}{\nonumber \\}
\newcommand{\bra}[1]{\mbox{$\langle \, {#1}\, |$}}
\newcommand{\ket}[1]{\mbox{$| \, {#1}\, \rangle$}}
\newcommand{\exval}[1]{\mbox{$\langle \, {#1}\, \rangle$}}

\def\bbbc{{\mathchoice {\setbox0=\hbox{$\displaystyle\rm C$}\hbox{\hbox
to0pt{\kern0.4\wd0\vrule height0.9\ht0\hss}\box0}}
{\setbox0=\hbox{$\textstyle\rm C$}\hbox{\hbox
to0pt{\kern0.4\wd0\vrule height0.9\ht0\hss}\box0}}
{\setbox0=\hbox{$\scriptstyle\rm C$}\hbox{\hbox
to0pt{\kern0.4\wd0\vrule height0.9\ht0\hss}\box0}}
{\setbox0=\hbox{$\scriptscriptstyle\rm C$}\hbox{\hbox
to0pt{\kern0.4\wd0\vrule height0.9\ht0\hss}\box0}}}}

\def\bbbq{{\mathchoice {\setbox0=\hbox{$\displaystyle\rm
Q$}\hbox{\raise 0.15\ht0\hbox to0pt{\kern0.4\wd0\vrule 
height0.8\ht0\hss}\box0}}
{\setbox0=\hbox{$\textstyle\rm Q$}\hbox{\raise
0.15\ht0\hbox to0pt{\kern0.4\wd0\vrule height0.8\ht0\hss}\box0}}
{\setbox0=\hbox{$\scriptstyle\rm Q$}\hbox{\raise
0.15\ht0\hbox to0pt{\kern0.4\wd0\vrule height0.7\ht0\hss}\box0}}
{\setbox0=\hbox{$\scriptscriptstyle\rm Q$}\hbox{\raise
0.15\ht0\hbox to0pt{\kern0.4\wd0\vrule height0.7\ht0\hss}\box0}}}}

\def\bbbt{{\mathchoice {\setbox0=\hbox{$\displaystyle\rm
T$}\hbox{\hbox to0pt{\kern0.3\wd0\vrule height0.9\ht0\hss}\box0}}
{\setbox0=\hbox{$\textstyle\rm T$}\hbox{\hbox
to0pt{\kern0.3\wd0\vrule height0.9\ht0\hss}\box0}}
{\setbox0=\hbox{$\scriptstyle\rm T$}\hbox{\hbox
to0pt{\kern0.3\wd0\vrule height0.9\ht0\hss}\box0}}
{\setbox0=\hbox{$\scriptscriptstyle\rm T$}\hbox{\hbox
to0pt{\kern0.3\wd0\vrule height0.9\ht0\hss}\box0}}}}

\def\bbbs{{\mathchoice
{\setbox0=\hbox{$\displaystyle     \rm S$}\hbox{\raise0.5\ht0\hbox
to0pt{\kern0.35\wd0\vrule height0.45\ht0\hss}\hbox
to0pt{\kern0.55\wd0\vrule height0.5\ht0\hss}\box0}}
{\setbox0=\hbox{$\textstyle        \rm S$}\hbox{\raise0.5\ht0\hbox
to0pt{\kern0.35\wd0\vrule height0.45\ht0\hss}\hbox
to0pt{\kern0.55\wd0\vrule height0.5\ht0\hss}\box0}}
{\setbox0=\hbox{$\scriptstyle      \rm S$}\hbox{\raise0.5\ht0\hbox
to0pt{\kern0.35\wd0\vrule height0.45\ht0\hss}\raise0.05\ht0\hbox
to0pt{\kern0.5\wd0\vrule height0.45\ht0\hss}\box0}}
{\setbox0=\hbox{$\scriptscriptstyle\rm S$}\hbox{\raise0.5\ht0\hbox
to0pt{\kern0.4\wd0\vrule height0.45\ht0\hss}\raise0.05\ht0\hbox
to0pt{\kern0.55\wd0\vrule height0.45\ht0\hss}\box0}}}}

\def\bbbz{{\mathchoice {\hbox{$\sf\textstyle Z\kern-0.4em Z$}}
{\hbox{$\sf\textstyle Z\kern-0.4em Z$}}
{\hbox{$\sf\scriptstyle Z\kern-0.3em Z$}}
{\hbox{$\sf\scriptscriptstyle Z\kern-0.2em Z$}}}}

\def\gsim {\mbox{\hbox{ \lower-.6ex\hbox{$>$}
\kern-1.12em \lower.5ex\hbox{$\sim$}\kern+.35em}}}
\def\lsim {\mbox{\hbox{ \lower-.6ex\hbox{$<$}
\kern-1.12em \lower.5ex\hbox{$\sim$}\kern+.35em}}}

\begin{document}

\begin{center}          
{\large                       
{\bf Nonequilibrium field-induced phase separation in single-file diffusion}
}\\[3cm]
{\large {
Fatemeh Tabatabaei\footnote{f.tabatabaei@fz-juelich.de}$^{\dag}$ and Gunter M. Sch\"utz\footnote{g.schuetz@fz-juelich.de}$^{\dag\ddag}$ }}\\[8mm]
{\em \dag Institut f\"ur Festk\"orperforschung, Forschungszentrum J\"ulich,\\
52425 J\"ulich, Germany}\\[2mm]
{\em \ddag Interdisziplin\"ares Zentrum f\"ur komplexe Systeme \\
University of Bonn, Germany}\\[4mm]

\vspace{1cm}                
\begin{minipage}{13cm}{
\baselineskip 0.3in
Using an analytically tractable lattice model for reaction-diffusion 
processes of hard-core 
particles we demonstrate that under nonequilibrium conditions 
phase 
coexistence may arise even if the system is effectively one-dimensional 
as e.g. in the channel system of some zeolites or in artificial optical 
lattices. In our model involving two species of particles a steady-state 
particle current is maintained by a density gradient between the channel 
boundaries and by the influence of an external driving force. This leads to the 
development of a fluctuating but always microscopically sharp interface 
between two domains of different densities which are fixed by the boundary 
chemical potentials. The internal structure of the interface 
becomes very simple for strong driving force. 
We calculate the drift velocity and diffusion
coefficient of the interface in terms of the
microscopic model parameters.\\[4mm]
}
\end{minipage} \end{center} 

\newpage
\baselineskip 0.3in
\section{Introduction}
\label{intro}
Diffusion of particles in long and narrow channels has a long history
of theoretical investigation and has recently become a focus also of 
experimental interest e.g. in the study of molecular diffusion in 
zeolites \cite{Kukl96}, diffusion of colloidal particles in confined
geometry \cite{Wei00} or optical lattices \cite{Lutz04} and granular
diffusion \cite{Coup06}. In many such channel systems the particles
cannot pass each other. This mutual blocking phenomenon is known as 
single-file effect and leads to subdiffusive behaviour \cite{Harr65,vanB83}. 
Among other things the single-file effect is responsible for
low reaction effectivity in microporous catalysts \cite{Brza05} and is 
thus of technical importance in chemical engineering. The single-file 
effect occurs also in biological systems, examples being the motion of 
ribosomes along the m-RNA during protein synthesis \cite{Schu97,Basu06} or 
transport by molecular motors along microtubuli or actin filaments 
\cite{Nish05}. In single-file systems the longitudinal motion 
is the most important dynamical mode and makes such processes 
amenable to treatment by one-dimensional models \cite{Schu03,Schu05}.

Low-dimensional diffusive particle systems are of great interest also from
a thermodynamic point of view. In open boundary systems, {\it kept far from 
equilibrium} by maintaining a steady state particle current, various 
unexpected kinds of 
critical phenomena have been discovered in recent years, including 
boundary-induced phase transitions, phase separation and spontaneous 
symmetry breaking, see \cite{Schu03,Schu05} and references therein 
for a review. These finite-temperature critical phenomena have no 
counterpart in thermal equilibrium since in one-dimensional 
systems with short range interactions there is no mechanism that could 
prevent the  creation and growth of an island of the minority phase inside a 
domain of the majority phase. Therefore it is not possible to have
a phase-separated equilibrium state with
a stable and microscopically sharp interface between two fluctuating 
domains characterized by different values of the order parameter. 

Most of these nonequilibrium critical phenomena are not yet well-understood.
Given the interesting diffusion properties as well as the potential
for applications to catalytic reactions it would thus be interesting 
to explore critical phenomena in low-dimensional reaction-diffusion 
systems in more detail. Specifically, in this paper on 
one-dimensional reaction-diffusion systems we would like to 
investigate the existence and microscopic properties of interfaces
between coexisting nonequilibrium domains which are macroscopically 
different.

In order to set the stage and sharpen the question we begin with
some remarks of general nature and mention some results relevant
to our approach.
Systems of diffusing and reacting particles are usually described
macroscopically by hydrodynamic equations for coarse-grained quantities like
the particle density \cite{Fife79}. The density then represents the local 
order 
parameter specifying the spatial evolution of the
macroscopic state of the system. Such equations are usually
proposed on a phenomenological basis, 
paradigmatic examples being
the Burgers equation for driven diffusive systems with particle conservation
\cite{Burg74} or the Fisher
equation for reactive  systems without conservation law \cite{Fish37,Lebo88}.
These
equations are in general non-linear and exhibit shocks in some cases. This
means that the solution of the macroscopic equations may develop a
discontinuity even if the initial particle density is smooth. This means
that in these systems phase separation may occur. The
shock represents the interface between the two thermodynamically distinct 
phases.

This hydrodynamic description of phase separation is, however,
not fully satisfactory. It provides no insight into the microscopic
origin of the phenomenon, and it gives no information about the
internal structure of the shock. It could very well happen that
in a particle system described on hydrodynamic (Eulerian) time scale 
by an equation which has shock solutions no corresponding
microscopic discontinuity would
be observable on 
less coarse-grained space or time scales which are experimentally
relevant particularly for the quasi one-dimensional systems referred to
above.
In order to understand the structure of shocks and  the emergence of such
nonlinear behaviour from the microscopic laws that govern the stochastic
motion and interaction of
particles it is therefore necessary to {\it derive} the macroscopic equations
from the microscopic dynamics rather than postulating them on phenomenological
grounds.

Carrying out this programme starting from Newton's equation of motion
constitutes a rather difficult problem. However,
a substantial body of results of this nature has been obtained for specific
one-dimensional stochastic lattice models
\cite{Spoh91,Kipn99}, the best-studied example being the asymmetric
simple exclusion process (ASEP) \cite{Ligg99,Schu00}. In this
basic model for a
driven diffusive system each site $k$ on the infinite integer lattice $\bbbz$ 
is either empty ($n_k=0$) or occupied by at most one particle ($n_k=1$). A 
particle on site $k$ hops randomly to the site $k+1$ with rate $D_r$ and to 
the site $k-1$ with rate $D_l$, but only if the target site is empty.
Otherwise the attempted move is rejected. The jumps occur independently in
continuous time with an exponential waiting time distribution. For a single
particle this is a biased random walk which on large scales
describes Brownian motion driven
by a constant external force. The exclusion rule mimics a short-ranged
hard-core interaction potential between particles.
 In the hydrodynamic limit the
system is described by the Burgers equation which exhibits
shocks. Such a shock discontinuity may be viewed as the interface between
stationary domains of different densities.

Moreover, there are a number of exact results about shocks in lattice gas models for
driven diffusive systems 
\cite{Ferr91,Derr97,Derr98,Pigo00,Bala01,Beli02,Kreb03,Rako04,Bala04,Jafa05,Arab06}, in
reaction-diffusion systems
\cite{Kreb03,Arab06,Doer91,Hinr96,benA98,Paes04} (where shocks appear as
Fisher waves on the macroscopic scale) and in spin-flip systems 
\cite{Kreb03,Arab06,Paes04} where shocks
correspond to domain walls \cite{Glau63}. It has emerged that in all these
models
the macroscopic shock discontinuity originates from a microscopically
sharp increase of the local particle density, i.e., a decrease of the
mean distance between particles that
can be observed on the scale of a few lattice units (which typically
represents the size of particles). The discontinuity itself performs
a biased random motion with a constant mean speed
and diffusive mean square displacement. The existence, structure, 
and dynamical properties of microscopically 
sharp shocks in lattice models for reaction-diffusion systems 
are the issues on which the present work focuses.

These results for the dynamical behaviour and microscopic properties
of shocks have been obtained for
infinite or periodic particle systems. In most physical applications, 
however, one has to study finite systems with
open boundaries where particles are injected and extracted.
This is crucial to take into account as -- in the absence of
equilibrium conditions --
the boundary conditions determine the bulk behavior of
driven systems, even to the extent that
boundary induced phase transitions between bulk states of different densities
occur \cite{Krug91,Schu93,Derr93}.
Qualitatively, the strong effect of boundary conditions on the bulk can be
attributed to the presence of steady-state currents which carry boundary
effects
into the bulk of the system. Quantitatively, exact results for the
steady state of the ASEP have helped to show that
part of the nonequilibrium phase diagram of driven diffusive systems with
open boundaries, viz. phase transitions of first order, can be understood
from the diffusive motion of shocks \cite{Kolo98,Popk99}, in analogy
to the Zel'dovich theory of equilibrium kinetics of first-order transitions.
In a series of recent papers \cite{Parm03,Popk03,Evan03,Rako03} these
considerations, originally formulated for conservative dynamics, have
been extended to non-conservative reaction-diffusion systems with open
boundaries. As in equilibrium, the nonequilibrium theory of boundary-induced
phase transitions requires the existence of shocks which are microscopically
sharp. Therefore, the study of the microscopic structure of shocks
in open systems is essential for understanding boundary-induced first
order transitions and the phase separation phenomena associated with
it.

After this survey we are finally in a position to precisely state
the objective of this work.
All the systems studied so far allow only for the presence
of a single species of particles. No exact
results have been reported so far for non-stationary travelling waves
in open two-component systems, i.e. where two diffusive particles species 
$A,B$ react with each other to form an inert reaction product
or undergo a cracking or coagulation reaction ($B \rightleftharpoons 2A$).
In order to address this question we adapt the strategy suggested in 
\cite{Kreb03} to two-component systems:
We take as initial distribution of particles
a shock distribution with given microscopic properties and look for
families of models for which the shock distribution evolves into a
linear combination of similar distributions with different
shock positions. Thus the information of the microscopic structure
of the shock that one has initially is preserved for all times.
Remarkably it will transpire that such families of reaction-diffusion
systems exist for strong
external field that drives the particles and keeps them in
a nonequilibrium state. We remark that in a similar treatment for
a different family of two-component processes we have found such a 
phenomenon at some specific finite driving strength \cite{Taba06a}.

The paper is organized as follows: In the following section we define
the class of models that we consider and we also define shock measures
for these systems. In Sec.~3 we determine the families of models
with travelling shock  solutions on the finite lattice.
In Sec.~4 we summarize our results and draw some conclusions.
Some mathematical details of the calculations are given in
the appendices.

\section{Stochastic reaction-diffusion processes}
\label{Sec2}

\subsection{Three-states lattice gas models}

In order to keep the physics that lead to phase-separated nonequilibrium 
states
as transparent as possible we study the simplest possible setting for a 
stochastic
two-component reaction-diffusion process. We consider a lattice gas model
defined on a lattice with $L$ sites.
The state of the system at any given time is described by a set of ``occupation 
numbers'' $\underline{n}={n_1,\dots,n_L}$ where $n_k=0,1,2$ is the local 
occupation number
at site $k$. These occupation numbers are abstract objects and serve as 
mathematical labels for three possible local states of each lattice site. 

The bulk stochastic dynamics
are defined by nearest neighbor transitions between the occupation variables
which occur independently and randomly in continuous time after an 
exponentially distributed waiting time. The mean 
$\tau(n_k',n_{k+1}';n_k,n_{k+1})$ of this waiting time depends on the 
transition $(n_k,n_{k+1}) \to (n_k',n_{k+1}')$. 
For later convenience we introduce an integer label 
\bel{label}
i=3n_k+n_{k+1}+1
\ee
in the range $1 \leq i \leq 9$ for the occupation variables on two 
neighboring sites $k$ and $k+1$. The inverse mean transition
times are the transition rates $w_{ij}$. Here $i=3n_k'+n_{k+1}'+1$ 
labels the target configuration and $j$ is the respective label 
of the initial configuration $(n_k,n_{k+1})$. We assume the bulk dynamics
to be spatially homogeneous. The transition rates
then do not explicitly depend on the site $k$. 

We require a single local conservation law where some linear 
function $C(n)$ of the local occupation numbers is conserved under the 
transitions \cite{Taba06a}. It is straightforward to 
check that this allows for 10 nonvanishing rates $w_{ij}$.
The physical interpretation of this conservation law as charge-, mass-, or particle conservation respectively depends on the physical interpretation of the occupation 
numbers $n_k$ and will be given below. We present the following three families of 
models which are mathematically equivalent, but have rather different physical interpretations.

\subsubsection{Diffusion without exclusion}

In its most obvious interpretation the abstract occupation number $n$
represents the number of particles on a given site. Requiring particle conservation where $C(n)=n$ 
allows for 10 hopping processes with rates given as follows:
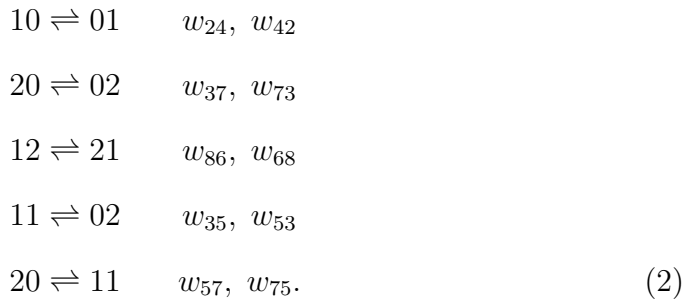
\begin{eqnarray}                           
  10 \rightleftharpoons 01
   \quad  & w_{24}, \; w_{42} \nonumber \\
  20 \rightleftharpoons 02
   \quad  & w_{37}, \; w_{73} \nonumber \\
  12 \rightleftharpoons 21
   \quad  & w_{86}, \; w_{68} \nonumber \\
  11 \rightleftharpoons 02
   \quad  & w_{35}, \; w_{53} \nonumber \\
  20 \rightleftharpoons 11
   \quad  & w_{57}, \; w_{75}.   \label{particleconserved}
\end{eqnarray}
Here there is no distinction between different particles, only the
total number is recorded. Physically this process describes diffusion of a 
single species of particles in a pore
system large enough to accommodate two particles in each pore.
Thus the three states do not describe a two-component, single-file
particle system, but a one-component system where particles can pass
each other. This makes this process different from the previously studied
two-state single-component systems which describe single-file diffusion 
\cite{Pigo00,Beli02,Kreb03}. For definiteness
we shall focus in this paper 
on two-component reaction-diffusion systems and hence not make use of this one-component 
interpretation of the three local states.

\subsubsection{Two-species annihilation $A+B\protect\rightleftharpoons0$}

We define 
\bel{chargedefinition}
C(n) = 1-n
\ee
as charge associated with the state $n$ of a lattice. The ``occupation number'' 
therefore denotes an internal degree of freedom in a single-file particle
system. The value $n=0$ 
corresponds to a positively charged particle (denoted as type $A$), $n=1$ 
corresponds to a vacant site (denoted $0$), and $n=2$ corresponds to occupation 
by a negatively charged particle (denoted as type $B$). As conservation law
we require charge conservation, or, equivalently, conservation of the
difference of particle numbers (of positively and negatively charged particles). 

Since this process is mathematically equivalent to the particle conserving
process (\ref{particleconserved})
the dynamics of the process can be represented by the following ten transitions
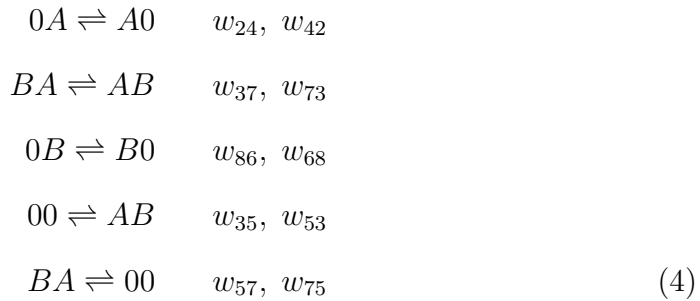
\begin{eqnarray}
  0A \rightleftharpoons A0
   \quad  & w_{24}, \; w_{42} \nonumber \\
  BA \rightleftharpoons AB
   \quad  & w_{37}, \; w_{73} \nonumber \\
  0B \rightleftharpoons B0
   \quad  & w_{86}, \; w_{68} \nonumber \\
  00 \rightleftharpoons AB
   \quad  & w_{35}, \; w_{53} \nonumber \\
  BA \rightleftharpoons 00
   \quad  & w_{57}, \; w_{75}  \label{chargconserved} 
\end{eqnarray}
This is the well-studied two-component creation/annihilation process,
see \cite{Priv97} for a review of some important properties and experimental
significance of
the one-dimensional pure annihilation case. The main results of this
paper are given in terms of this process.

\subsubsection{Cracking $B\rightarrow2A$}

One may switch the role of $A$ and $0$. 
The ``occupation number'' $n=0$ then corresponds to a vacant site $0$,
$n=1$ corresponds to a particle of type $A$, 
and $n=2$ corresponds to occupation by a particle of type $B$.
We drop the assignment of charges to particles and instead 
introduce
\bel{massdefintion}
C(n)=n=:M
\ee
as mass of the particles (in suitable units). 
$A$-particles thus have mass 1 and $B$-particles
to have mass 2; the conservation law 
describes mass conservation. Under this mapping the process
(\ref{chargconserved}) read
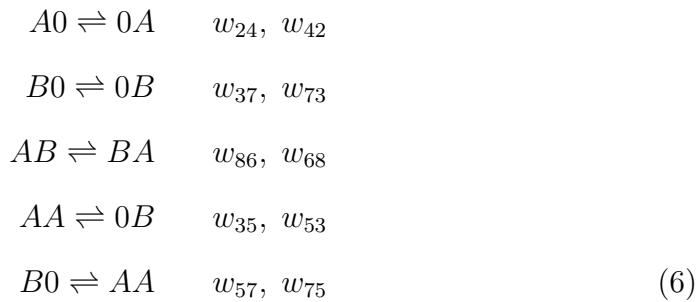
\begin{eqnarray}
  A0 \rightleftharpoons 0A
   \quad  & w_{24}, \; w_{42} \nonumber \\
  B0 \rightleftharpoons 0B
   \quad  & w_{37}, \; w_{73} \nonumber \\
  AB \rightleftharpoons BA
   \quad  & w_{86}, \; w_{68} \nonumber \\
  AA \rightleftharpoons 0B
   \quad  & w_{35}, \; w_{53} \nonumber \\
  B0 \rightleftharpoons AA
   \quad  & w_{57}, \; w_{75}  \label{massconserved} 
\end{eqnarray}
The last two reactions corresponds to cracking of a molecule $B$ 
with mass 2 into
two identical parts $A$ (mass 1 each), with coagulation
as reversed process. The third process in this list is a 
recombination reaction between neighboring reactands.

\subsubsection{Boundary conditions and continuity equation}

At the boundary sites $1,L$ we assume the system to be connected to
some external reservoir with which the system can exchange
particles. For definiteness we consider here and below charge
conservation. The corresponding processes for mass conservation
are obtained by changing $A\leftrightarrow 0$.

For injection and extraction of particles at the left boundary we introduce 
the rates :
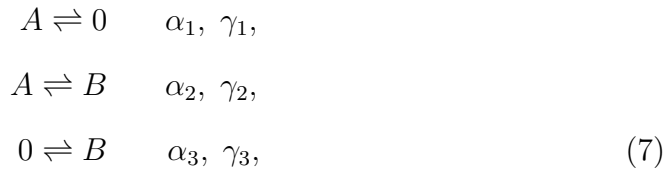
\begin{eqnarray}
  A \rightleftharpoons 0
   \quad & \alpha_1, \; \gamma_1, \nonumber \\
  A\rightleftharpoons B 
  \quad & \alpha_2,\; \gamma_2,  \nonumber \\   
 0 \rightleftharpoons B
   \quad & \alpha_3, \; \gamma_3,    
\end{eqnarray}
and for the right boundary 
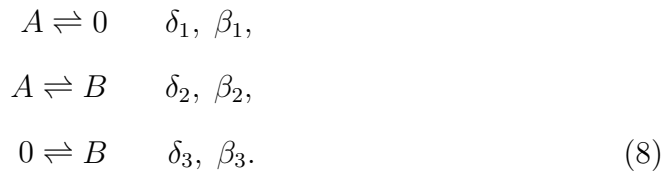
\begin{eqnarray}
  A \rightleftharpoons 0
   \quad & \delta_1, \; \beta_1, \nonumber \\
  A\rightleftharpoons B 
  \quad & \delta_2,\; \beta_2,  \nonumber \\   
 0 \rightleftharpoons B
   \quad & \delta_3, \; \beta_3.   
\end{eqnarray}

Here and below the left rate refers to the process going from
left to right, while the right rate is for the reversed process.
The boundary rates are a further set of model parameters.
Below we define them such that they are parametrized by 2 independent
boundary chemical potentials which fix boundary densities for the
conserved order parameter.

The presence of the bulk conservation law implies a lattice continuity equation
\bel{continuity}
\frac{d}{dt} C_k = j_{k-1} - j_k                              
\ee
for the expectation $C_k = \exval{C(n_k)}$. This quantity plays the role
of a conserved local order parameter. The quantity $j_k$ is the current
associated with the conservation law. It is given by the expectation
of some combination of local occupation numbers, depending on the model under
investigation, see below. Since we do not study here periodic systems
we do not require the boundary sites where the system is connected to
the reservoir to respect the conservation law. The quantities $j_0$, $j_L$
entering the continuity equation for $k=1$ and $k=L$ respectively
are source terms resulting from the 
reservoirs. They are functions of the reservoir densities. 
The lattice continuity equation is the starting point for a 
coarse-grained hydrodynamic description of the time evolution of the 
local order parameter.

\subsection{Master equation}

The time evolution defined above can be written in terms of a continuous-time 
master equation for the
probability vector
\begin{equation}
\ket{P(t)} = \sum_{\underline{n}} P(n_1, \cdots, n_L;t)\ket{\underline{n}},
\end{equation}
where $P(n_1, \cdots, n_L;t)$ is the distribution for the probability of 
finding particles at sites 1 to $L$ and $\ket{\underline{n}}$ is the basis 
vector in 
the space of configurations \cite{Schu00}. The probability vector is 
normalized such that 
$\langle s | P \rangle =1$ with the summation vector 
$\bra{s}=\sum_{\underline{n}}\bra{\underline{n}} $. 
The time evolution is generated by the stochastic Hamiltonian 
$H$ whose offdiagonal matrix elements $H_{\underline{n},\underline{n}'}$ are 
the negative transition rates between configurations. As required by 
conservation of probability, the diagonal elements 
are the negative sum of transition rates in the respective column.

Therefore the master equation is now  
described by the Schr\"odinger equation in imaginary time:
\begin{equation}
\frac{d}{dt}\ket{P(t)}= -H \ket{P(t)}.
\end{equation}
with the formal solution
\bel{2-12}
\ket{P(t)} = \mbox{e}^{-Ht}\ket{P(0)}.
\ee
Since only nearest-neighbour interactions are included, 
the quantum Hamiltonian $H$ defined above has the structure
\be      
H = b_1+\sum_{k=1}^{L-1} h_{k,k+1} + b_L.\label{eq:r30}
\ee
Here $b_1$ and $b_L$ are the boundary matrices:
\be
b_1 = - \left( \ba{cccc}
 -(\alpha_1+\alpha_2) & \gamma_1 & \gamma_2  \\
\alpha_1 & -(\gamma_1+\alpha_3) & \gamma_3  \\
\alpha_2 & \alpha_3 & -(\gamma_2+\gamma_3) \ea
\right)_1,\label{eq:r28}
\ee 
\vspace{1cm}
\be
b_L = - \left( \ba{cccc}
 -(\delta_1+\delta_2) & \beta_1 & \beta_2  \\
\delta_1 & -(\beta_1+\delta_3) & \beta_3  \\
\delta_2 & \beta_3 & -(\beta_2+\beta_3) \ea
\right)_L\label{eq:r29}.
\ee
The local bulk transition matrix $h_{k,k+1}$ with offdiagonal matrix
elements $-w_{ij}$ acts non-trivially only on sites 
$k$ and $k+1$. Below we give $h_{k,k+1}$ explicitly.

\subsection{Nonequilibrium steady states}

We stress that our family of models is defined in terms of transition rates,
not in terms of an internal energy $E(\underline{n})$ that would determine the 
stationary distribution of the process as equilibrium distribution $P^\ast(\underline{n}) 
\propto \exp{(-\beta E(\underline{n}))}$. Instead, the stationary
distribution is an {\it a priori} unknown and in general complicated function 
of 
the transition rates. It does not in general satisfy detailed balance
and thus represents a nonequilibrium steady state.
In order to be able to carry out explicit computations we 
restrict ourselves to systems such that the stationary
distribution of the stochastic dynamics factorizes, i.e., one has
a product measure without correlations
between the occupation numbers at different sites.

Requiring the existence of a stationary product measure imposes constraints 
both on the boundary rates and on the bulk rates. Physically, the 
constraints on the boundary rates essentially means that the
chemical potentials in the two reservoirs are equal, allowing
the bulk to relax into a current-carrying stationary state
with a chemical potential determined by the reservoirs. In this case the
origin of the current is not a gradient in the external chemical
potential of the reservoirs, but a constant bulk driving force.
The conditions on the bulk rates have a less transparent and model-dependent 
physical interpretation. Once these conditions are determined the model is
fully defined and its stationary distribution is given for equal 
chemical potentials in the reservoir.

In the quantum Hamiltonian formalism introduced above a
product measure is given by a tensor product
\begin{equation}
\ket{P}=|P_1)\otimes |P_2)\otimes ... \otimes |P_L).
\end{equation}
Here the three-component single-site probability vectors $|P_k)$ has
as its components the probabilities $P(n_k)$ of finding state $n$ at site
$k$. In the stationary distribution these probabilities are
position-independent, $|P_k) \equiv |P)$, and the stationary
probability vector thus has the form
\begin{equation}
\ket{P^*}=|P)^{\otimes L}.
\end{equation}

By definition of stationarity the stationary probability vector satisfies 
the eigenvalue equation
\begin{equation}
H\ket{P^*}=0 \label{eq:r1}.
\end{equation}
We shall parametrize the one-site marginals $P(n_k)$ by a generalized fugacity $z$
associated
with the conserved quantity and an interaction parameter determined by the
transition rates, see below. In formal analogy to equilibrium systems we shall
refer to the logarithm of the fugacity as chemical potential.

\subsection{Initial conditions}

The objective of this paper is the analysis of the family of models
which is defined by having a stationary product measure if the chemical potentials in the 
reservoir are equal. However, as physical boundary conditions to be studied we envisage {\it different} chemical potentials in the reservoirs. The
product measure is then no longer stationary and the questions arises
what new properties the stationary distribution exhibits and how
the system relaxes to its stationary distribution. Indeed,
in order to avoid misunderstanding we stress that the product
requirement on the stationary distribution with {\it equal} reservoir
chemical potentials
does not imply the absence of correlations during the time
evolution of the more general open system with different reservoir
chemical potentials.

Specifically, we 
prepare the system initially in a state described by a (nonstationary) shock
measure of the form
\begin{equation}
\ket{k}=|P_1)^{\otimes k} \otimes |P_2)^{\otimes L-k}.
\end{equation}
These shock measures have single-site probabilities given by $|P_1)$ in 
the left chain segment up to site $k$ (chosen to match the chemical
potential of the left reservoir) 
and single-site probabilities given by $|P_2)$ in the remaining
chain segment from site $L-k$ up to site $L$ (chosen to match
the chemical potential at the right reservoir). 

Such a shock measure 
defines fully the internal structure of the shock.  Since there 
are no correlations in a shock measure one may regard the lattice unit 
as the intrinsic 
shock width. A typical configuration has a sharp decrease of the mean 
interparticle
distance across the lattice point $k$. In the course of time
the measure $\ket{P(t)} = \exp(-Ht) \ket{k}$ changes and it is interesting
to investigate this time evolution.
For the
models studied below $\ket{P(t)}$ is computed explicitly and allows for
a detailed explicit calculation of all correlations that develop with
time.

\section{Field-induced phase separation}
\label{Sec3}

\subsection{Stationary distribution}

Following the ideas outlined above we first search for
stationary product solutions of the model with spatially 
constant single-site probabilities.
By choosing the basis of three states as follows
\begin{equation}
|A)=\left( \ba{c} 1 \\ 0\\0 \ea \right),\hspace{5mm}
|0)=\left( \ba{c} 0 \\ 1 \\ 0 \ea \right),\hspace{5mm}
|B)=\left( \ba{c} 0 \\ 0 \\ 1 \ea \right),
\end{equation}
one can conveniently write the product measure for the periodic model in 
terms of a generalized fugacity $z$ and arbitrary constant $r$
\begin{equation}\label{productmeasure}
\ket{P^*}=\frac{1}{\nu^L}\left( \ba{c} 1 \\ z \\ rz^2 \ea \right)^{\otimes L}
\end{equation}
Here
\begin{equation}
\nu=1+z+rz^2
\end{equation}
is the local ``partition function''. The quantity $r$ parametrizes
the density ratio of the two particle species, $\rho^B/\rho^A= rz^2$.
The fugacity $z$ is associated with the
conservation law, i.e., in a periodic system where the
charge is conserved $\ket{P^*}$ would be stationary 
for any value of $z$.
This probability measure is grand-canonical. The
charge $\sigma=\rho^A-\rho^B$ in this ensemble has mean
\be\label{charge}
\sigma = 1 - z \frac{d}{dz} \ln{\nu} = \frac{1-rz^2}{\nu}.
\ee
The corresponding canonical distributions with a definite value of the 
charge can be constructed in standard fashion, but we do not consider
them here since we are dealing with an open system where the bulk fugacity
is fixed by the generalized chemical potentials of the reservoirs. 
The nonconserved particle density $\rho=\rho^A+\rho^B$ in this ensemble
is given by 
\begin{equation}\label{density}
\rho=\frac{1+rz^2}{\nu}.
\end{equation}

The stationary distribution of the model is not known in full generality
and we have to determine constraints on the bulk rates such that the
product measure (\ref{productmeasure}) is stationary. The transition 
matrix $h_{k,k+1}$ for the bulk stochastic dynamics is given by
\[- h_{k,k+1}=
\]
\bel{2-16a}
\left( \ba{ccccccccc}
0 & 0 & 0 & 0 & 0 & 0 & 0 & 0 & 0\\
0 & -w_{42} & 0 & w_{24} & 0 & 0 & 0 & 0 & 0 \\ 
0 & 0 & -(w_{53}+w_{73}) & 0 &w_{35} &0 &w_{37} &0 &0 \\
0 & w_{42} & 0 & -w_{24} & 0 & 0 & 0 & 0 & 0\\
0 & 0 & w_{53}& 0& -(w_{35}+w_{75})& 0 & w_{57} & 0 & 0\\
0& 0 & 0 & 0 & 0  &-w_{86} & 0 & w_{68}& 0\\
0 & 0& w_{73}& 0& w_{75} &0 &-(w_{37}+w_{57})& 0& 0\\ 
0& 0& 0& 0& 0& w_{86} & 0 & -w_{68} & 0\\
0 &0 & 0  & 0 & 0 & 0 & 0 & 0 & 0 \\ \ea
\right)_{k,k+1}
\ee
and stationarity of the product measure implies
\begin{equation}
h_{k,k+1}\ket{P^*}=[F(\hat n^A_{k+1}-\hat n^A_{k}) +
F'(\hat n^B_{k+1} -\hat n^B_{k})]\ket{P^*}.\label{eq:r40}
\end{equation}
Here $F$ and $F'$ are arbitrary constants and $\hat n^A_k$ and 
$\hat n^B_k$ are number operators which take value 1 if there is
a particle of the respective species at site $k$ and zero otherwise,
i.e., $\rho^A = \exval{n^A_k}$ and $\rho^B = \exval{n^B_k}$
independently of $k$ due to homogeneity of the measure.

In order to satisfy the relation (\ref{eq:r40}) for systems with open 
boundaries 
we can write for $b_1$ and $b_L$, using another arbitrary constant $g$
\begin{equation}
b_1\ket{P^*}=(F\hat n^A_1+F'\hat n^B_1+g)\ket{P^*},
\end{equation}
\begin{equation}
b_L\ket{P^*}=-(F\hat n^A_L+F'\hat n^B_L+g)\ket{P^*}.
\end{equation}
As detailed in Appendix A one may solve the eigenvalue Eq.~(\ref{eq:r1}) 
and find $F$ and $F'$:
\bea
&F&=w_{24}-w_{42},\\
&F'&=w_{86}-w_{68}.
\eea
Therefore the bulk rates and densities satisfy two relations due to 
the eigenvalue equation (\ref{eq:r1}):
\be
 r=\frac{w_{35}+w_{75}}{w_{53}+w_{57}}\label{eq:r2},
\ee
\be
 w_{24}-w_{42}+w_{68}-w_{86}+w_{73}-w_{37}=\frac{w_{35}w_{57}-w_{53}w_{75}}
{w_{35}+w_{75}}.\label{eq:r3} 
\ee
The first equation (\ref{eq:r2}) expresses the constant $r$ in terms of the
reaction rates. The second equation (\ref{eq:r3}) is a constraint on the 
transition rates
which we impose on the model.

For the boundaries one needs to satisfy
\begin{equation}
g=\frac{1}{\nu}(F+F'rz^2).
\end{equation}      
This leaves two equations for the left boundary:
\begin{equation}
(w_{42}-w_{24})z(1+rz)-(w_{68}-w_{86})rz^2+ 
(\alpha_1+\alpha_2)\nu-\gamma_1z\nu-\gamma_2rz^2\nu=0,
\end{equation}      
\begin{equation}
(w_{68}-w_{86})rz^2(1+z)-(w_{42}-w_{24})rz^2-
\alpha_3z\nu+(\gamma_3+\gamma_2)rz^2\nu-\alpha_2\nu=0,
\end{equation}
and for the right boundary one has
\begin{equation}
(w_{42}-w_{24})z(1+rz)-(w_{68}-w_{86})rz^2-
(\delta_1+\delta_2)\nu+\beta_1\nu+\beta_2\mu=0,
\end{equation}
\begin{equation}
(w_{68}-w_{86})rz^2(1+z)-(w_{42}-w_{24})rz^2+
\delta_3z\nu-(\beta_3+\beta_2)rz^2\nu+\delta_2\nu=0.
\end{equation}
These equations relate the boundary rates to the fugacity and moreover
impose some constraints on the the boundary rates which are required
for a proper interpretation as boundary reservoirs with fixed chemical
potential.

We remark that the given choice of nonvanishing rates is only determined 
by the conservation law and requiring stationarity of the product measure. 
Many physical processes satisfy $PT$-invariance, 
i.e., the bulk dynamics should be symmetric under 
combined time reversal and space reflection. This physical input generalizes 
the equilibrium condition of detailed balance to allow for external driving 
forces. On the microscopic level of rates such driving forces
lead to a bias in the hopping rates. Well-known examples for models of this 
kind are exclusion processes satisfying pairwise balance \cite{Schu96}. 
As in the more general case discussed so far, the system is forced 
into a nonequilibrium steady state with a stationary current flowing in the 
system. Following \cite{Taba06a} we find that $PT$-invariance 
imposes the following further relations
\bea
w_{75}=rw_{53},\nn
w_{35}=rw_{57}.
\eea
In the calculations of the next section we do not make use of these
extra relations. We have merely listed them for possible applications
of our general results to specific $PT$-symmetric systems.
   
\subsection{Stationary Current and Hydrodynamics}

As remarked above the conservation law implies a lattice continuity equation
(\ref{continuity}) for the charge current.
To calculate the charge current we use the equation 
of motion for the expected local charge density
\begin{equation}
\frac{d}{dt} \sigma_k(t) =\frac{d}{dt}[\langle n^A_k\rangle-\langle 
n_k^B\rangle]= j_{k-1}-j_k.
\end{equation}
One finds for the expected local density of $A$-particles
\bea
\frac{d}{dt}\langle n^A_k\rangle=
&-w_{24}\langle n^0_{k-1} n^A_k \rangle +w_{42}
\langle n^A_{k-1} n^0_k \rangle-w_{37}\langle n^B_{k-1}n^A_k\rangle 
+w_{73}\langle n^A_{k-1}n^B_k\rangle \nn
&-w_{57}\langle n^B_{k-1}n^A_k\rangle + 
w_{75}\langle n^0_{k-1}n^0_k\rangle+
w_{24}\langle n^0_kn^A_{k+1} \rangle-w_{42}
\langle n^A_kn^0_{k+1} \rangle \nn
&+w_{37}\langle n^B_kn^A_{k+1} \rangle-
w_{73}\langle n^A_kn^B_{k+1}\rangle+
w_{35}\langle n^0_kn^0_{k+1}\rangle-w_{53}
\langle n^A_kn^B_{k+1}\rangle,\nn
\eea
and for $B$-particles
\bea
\frac{d}{dt}\langle n^B_k\rangle=&w_{37}\langle n^B_{k-1} n^A_k \rangle -
w_{73}\langle n^A_{k-1} n^B_k \rangle-w_{86}\langle n^0_{k-1}n^B_k\rangle+
w_{68}\langle n^B_{k-1}n^0_k\rangle\nn
&+w_{35}\langle n^0_{k-1}n^0_k\rangle-w_{53}\langle n^A_{k-1}n^B_k\rangle -
w_{37}\langle n^B_kn^A_{k+1} \rangle+w_{73}\langle n^A_kn^B_{k+1} \rangle \nn
&+w_{86}\langle n^0_kn^B_{k+1} \rangle-w_{68}\langle n^B_kn^0_{k+1}\rangle-
w_{57}\langle n^B_kn^A_{k+1}\rangle+w_{75}\langle n^0_kn^0_{k+1}\rangle \nn
\eea
This gives the charge current  
\bea
j_k=&-w_{24}\langle n^0_kn^A_{k+1}\rangle+w_{42}\langle n^A_kn^0_{k+1}\rangle-
2w_{37}\langle n^B_kn^A_{k+1}\rangle+2w_{73}\langle n^A_kn^B_{k+1}\rangle\nn
&-w_{68}\langle n^B_kn^0_{k+1}\rangle+w_{86}\langle n^0_kn^B_{k+1}\rangle-
w_{35}\langle n^0_kn^0_{k+1}\rangle+w_{53}\langle n^A_kn^B_{k+1}\rangle\nn
&-w_{57}\langle n^B_kn^A_{k+1}\rangle+w_{75}\langle n^0_kn^0_{k+1}\rangle .\nn
\eea  
In the steady state we can compute the current using the stationary 
distribution.
One finds
\begin{equation}
j^*=(-w_{24}+w_{42})\frac{z}{\nu^2}+[2(-w_{37}+w_{73})+w_{53}-w_{57}]
\frac{rz^2}{\nu^2}+(w_{86}-w_{68})\frac{rz^3}{\nu^2}+(w_{75}-w_{35})
\frac{z^2}{\nu^2},
\end{equation}
and by using (\ref{eq:r2}) and the stationary condition (\ref{eq:r3})  
\begin{equation}
j^*=\frac{1}{2}(w_{42}-w_{24})(\rho+\sigma)(1-\sigma)+\frac{1}{2}
(w_{86}-w_{68})(\rho-\sigma)(1+\sigma),\label{eq:r25}
\end{equation} 
where $\sigma$ and $\rho \equiv \langle n^A_k \rangle+\langle n^B_k \rangle$ 
are the 
stationary density of charges (\ref{charge}) and particles (\ref{density}) respectively.

Since the individual particle densities are not conserved the equations of
motion for the local densities take the form
\begin{equation}
\frac{d}{dt}\langle n^A_k\rangle= j^A_{k-1}- j^A_k+S_k,
\end{equation}
\begin{equation}
\frac{d}{dt}\langle n^B_k\rangle=j^B_{k-1}- j^B_k +S_k,
\end{equation}
with source term
\bea
S_k=&-\frac{1}{2}w_{57}(\langle n^B_{k-1} n^A_k\rangle+
\langle n^B_{k}n^A_{k+1}\rangle)+
\frac{1}{2}w_{75}(\langle n^0_{k-1} n^0_k \rangle+
\langle n^0_k n^0_{k+1}\rangle)\nn
&+\frac{1}{2}w_{35}(\langle n^0_{k-1}n^0_k \rangle+ 
\langle n^0_k n^0_{k+1} \rangle) -\frac{1}{2}w_{53}
(\langle n^A_{k-1} n^B_k\rangle+ \langle n^A_k n^B_{k+1}\rangle).\nn
\eea
The particle currents are given by
\bea
j^A_k=&-&w_{24}\langle n^0_kn^A_{k+1}\rangle+w_{42}
\langle n^A_kn^0_{k+1}\rangle-w_{37}\langle n^B_kn^A_{k+1}
\rangle+w_{73}\langle n^A_kn^B_{k+1}\rangle\nn
&-&\frac{1}{2}w_{35}\langle n^0_kn^0_{k+1}\rangle+\frac{1}{2}
w_{53}\langle n^A_kn^B_{k+1}\rangle-\frac{1}{2}w_{57}
\langle n^B_kn^A_{k+1}\rangle+\frac{1}{2}w_{75}\langle n^0_kn^0_{k+1}
\rangle,\nn
\eea
\bea
j^B_k=&&w_{68}\langle n^B_kn^0_{k+1}\rangle-w_{86}
\langle n^0_kn^B_{k+1}\rangle+w_{37}\langle n^B_kn^A_{k+1}
\rangle-w_{73}\langle n^A_kn^B_{k+1}\rangle\nn
&+&\frac{1}{2}w_{35}\langle n^0_kn^0_{k+1}\rangle-\frac{1}{2}
w_{53}\langle n^A_kn^B_{k+1}\rangle+\frac{1}{2}w_{57}
\langle n^B_kn^A_{k+1}\rangle-\frac{1}{2}w_{57}\langle n^0_kn^0_{k+1}
\rangle.\nn
\eea
The resulting charge current $j_k=j^A_k-j^B_k$ is studied above. 
One may introduce
also a particle current $\tilde j_k=j^A_k+j^B_k $ and write           
\begin{equation}
\frac{d}{dt} \rho_k(t)=\tilde j_{k-1}-\tilde j_k+2S_k.
\end{equation}

For a coarse-grained hydrodynamic description of the time-evolution of the 
system
we follow standard arguments \cite{Spoh91,Kipn99}. We pass to a continuum
description by making the lattice unit $a$ 
(which until now had been taken to be $a=1$) infinitesimal and we consider continuum space as $x=\frac{k}{L}$. 
The coarse-grained local observables $\sigma(x,t),\rho(x,t)$ in continuous 
space
are averaged over a large but finite lattice interval around the lattice point 
$x$
and therefore given by the expected local densities $\sigma_x(t),\rho_x(t)$.
We consider Eulerian scaling $t'=ta$ with rescaled macroscopic time $t'$.
In the continuum limit the two equations for $\sigma$ and $\rho$ 
then take the form (to 
leading order in the lattice constant $a$)
\begin{equation}
\partial_{t'}\sigma (x,t')=-\partial_x j(\sigma,\rho),
\end{equation}
\begin{equation}
\partial_{t'}\rho (x,t')=-\partial_x\tilde j(\sigma,\rho)+R(\sigma,\rho)/a 
+
\tilde R(\sigma,\rho),
\end{equation}
where because of local stationarity
\begin{equation}
R(\sigma,\rho)=-\frac{1}{2}(w_{57}+w_{53})(\rho+\sigma)
(\rho-\sigma)+
2(w_{75}+w_{35})(1-\rho)^2,
\end{equation}
\begin{equation}
\tilde R(\sigma,\rho)=(w_{57}-w_{53})\frac{1}{4}[(\rho+\sigma) 
\partial_x(\rho-\sigma)-(\rho-\sigma)\partial_x(\rho+\sigma)].
\end{equation}
The space-time dependence of $R$ and $\tilde R$ is implicit in
arguments $\sigma(x,t'),\rho(x,t')$.

In this limit, when time and space are large, the term contained
$R(\sigma,\rho)$ in the equation for $\rho$ becomes large enough 
to make the two other terms negligible.
Therefore $\rho(x,t)$ reaches its stationary state very fast, in agreement with 
the
notion that non-conserved local degrees of freedom have attained their
stationary values under hydrodynamic scaling. This implies that in the
stationary state
$R(\sigma,\rho)=0$, from which we obtain the stationary particle
density
\begin{equation}
(\rho^{*2}-\sigma^2)=4r(1-\rho^*)^2.\label{eq:r26}    
\end{equation}
for a given value of charge $\sigma$.
Therefore $\rho$ takes at any instant of (macroscopic) time a special value 
$\rho^*$ which is a function of $\sigma$.
The remaining slow dynamical mode is the charge, the evolution of
which is thus governed by the hydrodynamic equation
\bel{hydro1}
\partial_{t'}\sigma(x,t')=-\partial_x j_x(\rho^*,\sigma) 
= - \partial_\sigma j(\rho^*,\sigma) \partial_x \sigma(x,t').
\end{equation}
In the second equation $j(\rho^*,\sigma)$ is the stationary current
(\ref{eq:r25}). 
This evolution equation is a nonlinear partial
differential equation which can be solved by the method of
characteristics. Because of the nonlinearity the solution may 
develop shocks in the charge distribution and we now turn
to the investigation of these shocks on microscopic scale.

\subsection{Shock measures}

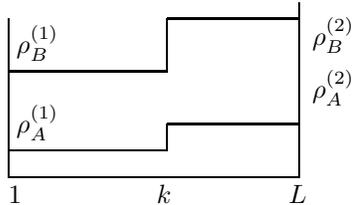
\begin{figure}[h]
\begin{center}
\begin{picture}(120,70)
\put(10,10){\line(0,1){60}}
\put(10,10){\line(1,0){110}}
\put(10,20){\line(1,0){60}}
\put(70,20){\line(0,1){10}}
\put(70,30){\line(1,0){50}}
\put(120,10){\line(0,1){66}}
\put(10,50){\line(1,0){60}}
\put(70,50){\line(0,1){20}}
\put(70,70){\line(1,0){50}}
\put(10,0){\footnotesize 1}
\put(66,0){\footnotesize $k$}
\put(116,0){\footnotesize $L$}
\put(125,40){\footnotesize $\rho_A^{(2)}$}
\put(13,27){\footnotesize $\rho_A^{(1)}$}
\put(125,60){\footnotesize $\rho_B^{(2)}$}
\put(13,57){\footnotesize $\rho_B^{(1)}$}
\end{picture}
\caption[profig]{Coarse grained density profiles of a shock measure with shock 
between 
sites $k,k+1$.}
\label{profig}
\end{center}
\end{figure}

We assume that the initial distribution of charges exhibits a
shock which on microscopic scale is represented by a shock measure 
(see Figure \ref{profig}). We represent a shock measure with a 
shock in the fugacities between sites $k$ and $k+1$ as
\begin{equation} 
\ket k=\frac{1}{\nu_1^k \nu_2^{L-k}}\left( \ba{c} 1 \\ z_1 
\\ rz_1^2 \ea \right)^{\otimes k}\otimes \left( \ba{c} 1 \\ z_2 
\\ rz_2^2 \ea \right)^{\otimes L-k}.
\end{equation} 
In this model with open boundary condition, the first (second) fugacity
matches the fugacity in left(right) boundary. 

Now we investigate the possibility that in analogy to the processes considered 
in 
\cite{Kreb03,Paes04,Rako04} the family of shock 
measures $\ket k$ is closed under the time 
evolution $t$. This means that the initial measure  evolves into a 
linear combination of shock measures after time $t$. This condition 
requires $H$ which generates the time evolution to satisfy the following
equation after an infinitesimal step
\begin{equation}
\frac {d}{dt} \ket k = d_1 \ket{k-1} +d_2 \ket{k+1}-(d_1+d_2)\ket k.
\label{eq:r4}
\end{equation}
We remark that this equation for the full particle distribution is 
mathematically 
equivalent to the evolution equation of a single-particle 
random walk with hopping rate $d_1$ to the left and $d_2$ to the right.
Thus, if (\ref{eq:r4}) can be satisfied, the shock in the initial distribution
remains microscopically preserved at all times, but its position performs a 
random walk with shock hopping rates $d_1$ to the left and $d_2$ to the right
respectively.

For further analysis we define
\begin{equation}
\tilde{h}_{i,i+1} \equiv h_{i,i+1} + F (n^A_i -n^A_{i+1}) + F' (n^B_i -n^B_{i+1}),\label{eq:hi}
\end{equation}
\begin{equation}
\tilde{b}_1\equiv b_1-Fn^A_1-F'n^B_1,\label{eq:b1}
\end{equation}
\begin{equation}
\tilde{b}_L\equiv b_L+Fn^A_L+F'n^B_L.\label{eq:bl}
\end{equation}
Using 
\begin{equation}
\tilde{h}_{i,i+1}\ket k=0 \hspace{4mm}  for \hspace{2mm} i\neq k,
\end{equation}
\begin{equation}
\tilde{b}_1\ket k= g_1\ket k, \hspace{5mm} \tilde{b}_L\ket k =-g_2 \ket k.
\end{equation}
with
\begin{equation}
g_1=-F\frac{1}{\nu_1}-F'\frac{rz_1^2}{\nu_1},\label{eq:r20}
\end{equation}
\begin{equation}
g_2=-F\frac{1}{\nu_2}-F'\frac{rz_2^2}{\nu_2},\label{eq:r21}
\end{equation}
yields
\begin{equation}
-H \ket k=-( \sum_i{\tilde{h}_{i,i+1}}+\tilde{b}_1+\tilde{b}_L)\ket k
=(-\tilde{h}_{k,k+1}-g_1+g_2)\ket k.
\end{equation}
Together with
(\ref{eq:r4}) we thus find
\begin{equation}
(-\tilde{h}_{k,k+1}+d_1+d_2-g_1+g_2)\ket k-d_1\ket{k-1} - d_2\ket{k+1}=0.
\label{randomwalkcondition}
\end{equation}
The quantities $g_{1,2}$ are obtained from the boundary conditions
(Appendix B).

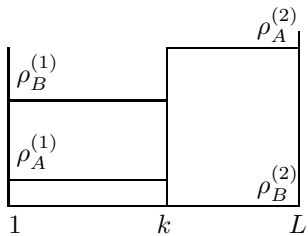
\begin{figure}[h]
\begin{center}
\begin{picture}(120,70)
\put(10,10){\line(0,1){60}}
\put(10,10){\line(1,0){110}}
\put(10,20){\line(1,0){60}}
\put(70,20){\line(0,-1){10}}
\put(70,10){\line(1,0){50}}
\put(120,10){\line(0,1){66}}
\put(10,50){\line(1,0){60}}
\put(70,50){\line(0,-1){40}}
\put(70,10){\line(1,0){10}}
\put(70,50){\line(0,1){20}}
\put(70,70){\line(1,0){50}}
\put(10,0){\footnotesize 1}
\put(66,0){\footnotesize $k$}
\put(116,0){\footnotesize $L$}
\put(13,27){\footnotesize $\rho_A^{(1)}$}
\put(104,15){\footnotesize $\rho_B^{(2)}$}
\put(13,57){\footnotesize $\rho_B^{(1)}$}
\put(104,75){\footnotesize $\rho_A^{(2)}$}
\end{picture}
\caption[profigg]{Density profile of a shock measure in the case $z_2=0$}
\label{profigg}
\end{center}
\end{figure}

This is a set 9 equations for the bulk rates. We have found a solution
(see Appendix B) with $w_{24}=0$. Putting this into the 9 equations 
(\ref{eq:r5})-(\ref{eq:r13})
one finds after some straightforward algebra 
\begin{equation}
d_2=z_2=0,
\end{equation}
\begin{equation}
d_1=\frac{S}{\nu_1}=\frac{w_{42}}{\nu_1},
\end{equation}
\begin{equation}
w_{57}=w_{37}=0,
\end{equation}
\begin{equation}             
w_{86}=w_{68}.
\end{equation}
In this model there is a strong driving force for the positive
particles that leads them to move only to the right as in the
totally asymmetric simple exclusion process. $z_2=0$ means that in 
the right branch of the shock the lattice is completely filled with positive 
particles (see Fig. \ref{profigg}). Hence incoming $A$-particles
which react with $B$ particles in the left branch of the shock hit
the pure $A$-domain where they stop because of the single-file
(exclusion) condition. The shock that separates the two domain
moves only to the left with rate $d_1$. Hence its mean velocity $v_s$
and diffusion coefficient $D_s$ are determined by the density and hopping
rate only of the $A$-particles in the left domain
\bel{shockparameters} 
v_s =2D_s = w_{42} \rho^A_1.
\ee
\vspace{1cm}
                
The interpretation of this result for the cracking process is
readily available by interchanging the role of positive
particles and vacancies. The right branch of the shock is the
empty lattice where no reactions are going on. The left branch is 
active. All particles are driven to the left so that the inactive
region grows diffusively with drift and fluctuations determined by
(\ref{shockparameters}) and $\rho^A_1$ replaced by the vacancy
density in the active domain.

We note that $PT$-invariance of the special model with $w_{24}=0$ leads to
\begin{equation}
w_{35}=0,
\end{equation}
and to the stationary state condition 
\begin{equation}
w_{42}=w_{73}+w_{53}.
\end{equation} 
The properties of the shock are not effected by $PT$-invariance.

\section{Conclusions}

We have studied the dynamics of a family of one-dimensional
driven two-component reaction-diffusion 
processes with open boundaries on microscopic {\it lattice scale} and derived a
hydrodynamic description on coarse grained Eulerian scale. This is
the first main result, see Eqs. (\ref{eq:r25}), (\ref{eq:r26}), (\ref{hydro1}). The hydrodynamic
equation is nonlinear and therefore admits shock solutions, corresponding to
phase-separated states of the system. This generalizes one-dimensional
field-induced phase separation that has been studied in some detail 
for lattice fluids in thermal equilibrium \cite{Robe84}.

As the second main result we have obtained for a subset of 
models with very strong driving force detailed knowledge about 
the microscopic structure of 
the shock. The transition between the two phases is maximally 
sharp on lattice scale 
and the shock position performs a biased random walk with drift velocity
and diffusion coefficient (\ref{shockparameters}).
Therefore, as
observed in other models, shocks behave
like collective single-particle excitations already on the lattice scale --
not only after coarse-graining where all the microscopic features of the
shock are lost. To our knowledge together with \cite{Taba06a} these are the first results of this
nature obtained for two-component reaction-diffusion processes.
The mapping to different models, in particular to the partial exclusion
process, suggests that this feature is not specific to single-file
diffusion. The driving force that is required to produce a maximally sharp
interface depends on the model under consideration \cite{Taba06a}. 

For a more general choice of rates the microscopic time evolution of the
shock structure is more complicated. By analogy with general considerations
of microscopic shock stability \cite{Schu00} and exact results for the
asymmetric simple exclusion process \cite{Derr97} it is natural to expect a 
microscopically sharp interface, but with some extended structure that depends
on the strength of the driving force. It would be very interesting to 
investigate experimentally in effectively one-dimensional
driven reaction-diffusion systems with a conservation law the existence of 
field-induced phase separation and force-dependence of the internal width 
of the domain boundary.

\subsection*{Acknowledgments}
FT would like to thank R.J. Harris for useful discussions.

\appendix

\section*{Appendix A: Stationarity condition}        
\begin{appendix}
\setcounter{equation}{0}
\renewcommand{\theequation}{A-\arabic{equation}}

Assuming product measure as stationary solution, we have 
\begin{equation}
\ket {P^*}=\frac{1}{\nu^L}\left( \ba{c} 1 \\ z \\ rz^2 \ea \right)^{\otimes 
L}.
\end{equation}
With (\ref{eq:hi})-(\ref{eq:bl}), where $n^A_i$ and $n^B_i$ are number operators
\begin{equation}
n^A_i=\left(\ba{ccc} 1&0&0\\0&0&0\\0&0&0\ea\right)_i, \hspace{1cm}n^B_i= 
\left(\ba{ccc} 0&0&0\\0&0&0\\0&0&1\ea\right)_i,
\end{equation}
and eigenvalue equation 
\begin{equation}
H\ket{P^*}=0,
\end{equation}
we write
\begin{equation}
\tilde{h}_{i,i+1}\ket{P^*}=(\tilde b_1+\tilde b_L)\ket{P^*}=0.
\end{equation}
$ \tilde h_{i,i+1} $ in terms of arbitrary constants $F$ and $F'$ is given by
\[ \tilde h_{i,i+1}=
\]
\bel{2-16c}
- \left( \ba{ccccccccc}
 0 & 0 & 0 & 0 & 0 & 0 & 0 & 0 & 0\\0 &-F-w_{42} & 0 & w_{24} & 0 & 0 & 0 & 0 
& 0 \\ 0 & 0 & -\Theta_1 & 0 &w_{35} &0 &w_{37} &0 &0 \\0 & w_{42} & 0 
&F-w_{24} & 0 & 0 & 0 & 0 & 0\\0 & 0 & w_{53}& 0& -(w_{35}+w_{75})& 0 & 
w_{57} & 0 & 0\\0& 0 & 0 & 0 & 0  &F'-w_{86} & 0 & w_{68}& 0\\0 & 0 & w_{73}& 
0& w_{75} &0 &-\Theta_2& 0& 0\\ 0& 0& 0& 0& 0&w_{86} & 0 &-F'-w_{68} & 0\\
0 &0 & 0  & 0 & 0 & 0 & 0 & 0 & 0 \\ \ea
\right)_{i,i+1},
\ee
where
\begin{equation}
\Theta_1=F-F'+w_{53}+w_{73},
\end{equation}
\begin{equation}
\Theta_2=-F+F'+w_{37}+w_{57}.
\end{equation}
Substituting $\tilde h_{i,i+1}$ in the Eq.~(A-7) yields 5 independent equations. One 
gets $F$ and $F'$ by solving following equations 
\begin{eqnarray}
(F+w_{42}-w_{24})z=0,\\
(-F'+w_{86}-w_{68})rz^2=0.
\end{eqnarray}
Hence $F$ and $F'$ are
\begin{equation}
F=w_{24}-w_{42},
\end{equation}
\begin{equation}
F'=w_{86}-w_{68}.
\end{equation} 
Three remained equations which have to be satisfied are
\begin{eqnarray}
(w_{37}-\Theta_1)rz^2+w_{35}z^2=0,\\
(w_{73}-\Theta_2)rz^2+w_{75}z^2=0,\\
(w_{53}+w_{57})rz^2-(w_{35}+w_{75})z^2=0.
\end{eqnarray}
From Eq.~(A-17) we obtain
\begin{equation}
r=\frac{w_{35}+w_{75}}{w_{53}+w_{57}}.
\end{equation}
Subtracting Eq.~(A-15) from Eq.~(A-16) yields second stationary state 
condition
\begin{equation}
w_{24}-w_{42}+w_{68}-w_{86}+w_{73}-w_{37}=\frac{w_{35}w_{57}-w_{53}w_{75}}
{w_{35}+w_{75}}, 
\end{equation}    
where the sum of (A-15) and (A-16) is already satisfied.
         
This model is assumed to have open boundaries, therefore $b_1$ and 
$b_L$ in terms of injection and extraction rates are 
given by Eq.~(\ref{eq:r28}) and 
Eq.~(\ref{eq:r29}).
For satisfying Eq.~(A-6) for the model with open boundaries one writes
\begin{equation}
b_1\ket{P^*}=(F\hat n^A_1+F'\hat n^B_1+g)\ket{P^*},
\end{equation}   
\begin{equation}
b_L\ket{P^*}=-(F\hat n^A_L+F'\hat n^B_L+g)\ket{P^*}.
\end{equation}
where $g$ is an arbitrary constant.
Eq.~(A-22) for the left boundary leads to three equations 
\bea
(\alpha_1+\alpha_2)-\gamma_1z-\gamma_2 rz^2=F+g,\\
-\alpha_1+(\gamma_1+\alpha_3)z-\gamma_3rz^2=gz,\\
-\alpha_2-\alpha_3z+(\gamma_2+\gamma_3)rz^2=(F'+g)rz^2.
\eea
We then obtain $g$ 
\bea
g&=&-\frac{F+F'rz^2}{\nu}\nn
&=&\gamma_1+\alpha_3-\frac{\alpha_1}{z}-rz\gamma_3.\nn
\eea
One also can obtain two conditions for boundary rates, which for the left one
\begin{equation}
(w_{42}-w_{24})z(1+rz)-(w_{68}-w_{86})rz^2+ 
(\alpha_1+\alpha_2)\nu-\gamma_1z\nu-\gamma_2rz^2\nu=0,
\end{equation}      
\begin{equation}
(w_{68}-w_{86})rz^2
(1+z)-(w_{42}-w_{24})rz^2-\alpha_3z\nu+(\gamma_3+\gamma_2)rz^2\nu-\alpha_2\nu=0,
\end{equation}
and for the right boundary 
\begin{equation}
(w_{42}-w_{24})z(1+rz)-(w_{68}-w_{86})rz^2-(\delta_1+\delta_2)\nu+\beta_1\nu+\beta_2\mu=0,
\end{equation}
\begin{equation}
(w_{68}-w_{86})\mu(1-\mu)-(w_{42}-w_{24})\nu\mu-(\beta_3+\beta_2)\mu+\gamma_3\lambda+\delta_2\nu=0.
\end{equation}         

\end{appendix}

\appendix
\section*{Appendix B: Random walk conditions for the shock}        
\begin{appendix}
\setcounter{equation}{0}
\renewcommand{\theequation}{B-\arabic{equation}}

Explicitly the equations (\ref{randomwalkcondition})
that solve the random-walk
condition for the shock are given by
\be
S-d_1\frac{\nu_1}{\nu_2}-d_2\frac{\nu_2}{\nu_1}=0,\label{eq:r5} 
\ee
\be
(S-w_{35}-w_{75})z_1z_2+w_{53}rz^2_2+w_{57}rz^2_1-d_1
\frac{\nu_1}{\nu_2}z_2^2-d_2\frac{\nu_2}{\nu_1}z_1^2=0,\label{eq:r6} 
\ee
\be
S-d_1\frac{z_2^2}{z_1^2}\frac{\nu_1}{\nu_2}-d_2
\frac{z_1^2}{z_2^2}\frac{\nu_2}{\nu_1}=0,\label{eq:r7} 
\ee
\be
(S-w_{24})z_2+w_{24}z_1-d_1z_2\frac{\nu_1}{\nu_2}-
d_2z_1\frac{\nu_2}{\nu_1}=0,\label{eq:r8} 
\ee
\be
(S-w_{42})z_1+w_{42}z_2-d_1z_2\frac{\nu_1}{\nu_2}-
d_2z_1\frac{\nu_2}{\nu_1}=0,\label{eq:r9} 
\ee
\be
(S-w_{68})z_2+w_{68}z_1-d_1\frac{z_2^2}{z_1}\frac{\nu_1}{\nu_2}-
d_2\frac{z_1^2}{z_2}\frac{\nu_1}{\nu_2}=0,\label{eq:r10} 
\ee
\be
(S-w_{86})z_1+w_{86}z_2-d_1\frac{z_2^2}{z_1}\frac{\nu_1}{\nu_2}-
d_2\frac{z_1^2}{z_2}\frac{\nu_1}{\nu_2}=0, \label{eq:r11}   
\ee
\be
(S-w_{37}-w_{53}-\Delta)rz^2_2+w_{35}z_1z_2+w_{37}rz_1^2-d_1rz_2^2
\frac{\nu_1}{\nu_2}-d_2rz_1^2\frac{\nu_2}{\nu_1}=0,\label{eq:r12} 
\ee
\be
(S-w_{73}-w_{57}+\Delta)rz_1^2+w_{75}z_1z_2+w_{73}rz_2^2-d_1rz_2^2
\frac{\nu_1}{\nu_2}-d_2rz_1^2\frac{\nu_2}{\nu_1}=0,\label{eq:r13}
\ee
where for compact notation we have introduced
\begin{equation}
S=d_1+d_2-g_1+g_2 ; \hspace{3mm}  
\Delta=w_{24}-w_{42}+w_{68}-w_{86}+w_{73}-w_{37}.
\end{equation}

These relations can be rewritten as 4 independent relations between the 
hopping rates and the fugacities
\begin{equation}
w_{24}=w_{68}\equiv p,\label{eq:r17}
\end{equation}
\begin{equation}
w_{42}=w_{86}\equiv q, \label{eq:r18}
\end{equation}
\begin{equation}
\frac{p}{q}=\frac{z_2^2}{z_1^2}\equiv X^2,\label{eq:r16}
\end{equation}
\begin{equation}
S=p+q.\label{eq:r19}
\end{equation}
and two equations for the shock hopping rates 
\begin{equation}
d_1=q\frac{\nu_2}{\nu_1}\label{eq:r14},
\end{equation}
\begin{equation}
d_2=p\frac{\nu_1}{\nu_2}\label{eq:r15}.
\end{equation}
To be more specific, solving Eq.~(\ref{eq:r5}) and (\ref{eq:r8})-(\ref{eq:r9}) 
yields Eq.~(\ref{eq:r14}) and Eq.~(\ref{eq:r15}) for $d_1$ and $d_2$, from 
these 
two and Eq.~(\ref{eq:r7}) and (\ref{eq:r10})-(\ref{eq:r11}), we obtain 
(\ref{eq:r16}), a relation between rates and densities, then using 
Eq.~(\ref{eq:r5}) with above results yields Eq.~(\ref{eq:r17}),
(\ref{eq:r18}) and (\ref{eq:r19}).

Using these 6 relations (\ref{eq:r17})-(\ref{eq:r15}), equations 
(\ref{eq:r12}), (\ref{eq:r13}) 
and Eq.~(\ref{eq:r6}) respectively lead to the following relations for 
the so far undetermined rates
\begin{equation}
(p-q)(1-\frac{w_{37}}{p})r+w_{35}(\sqrt{\frac{q}{p}}-1)=0,\label{eq:r22}
\end{equation}
\begin{equation}
(q-p)(1-\frac{w_{73}}{q})r+w_{75}(\sqrt{\frac{p}{q}}-1)=0,\label{eq:r23}
\end{equation}
\begin{equation}
(p+q)r^{-1}+w_{53}(\sqrt{\frac{p}{q}}-1)+w_{57}(\sqrt{\frac{q}{p}}-1)=0.
\label{eq:r24}
\end{equation}
Simplifying Eq.~(\ref{eq:r19}) by using Eqs.~(\ref{eq:r20})-(\ref{eq:r21}) for 
$g_1$ and $g_2$
 yields following more explicit relation between $r$ and $X$ 
\begin{equation}
r=\frac{X}{(1+X)^2}.
\end{equation}
This relation on $r$ together with Eqs.~(\ref{eq:r22})-(\ref{eq:r24}) 
and the stationary state equation (\ref{eq:r3}), 
implies that $X=0$. This solved by $p=z_2=0$.

The boundary equations (\ref{eq:r20})-(\ref{eq:r21}) lead to 
\bea
g_1&=&\frac{p-q}{\nu_1}(rz_1^2-1)\nn
&=&-\alpha_1\frac{1}{z_1}-\gamma_3 rz_1+(\gamma_1+\alpha_3),
\eea
\bea
g_2&=&\frac{p-q}{\nu_2}(rz_2^2-1)\nn
&=&\delta_1\frac{1}{z_2}+\beta_3 rz_2-(\beta_1+\delta_2),
\eea
for $g_1$ and $g_2$ respectively.

\end{appendix}

\newpage

\bibliographystyle{unsrt}

\begin{thebibliography}{99}

\bibitem{Kukl96}
V. Kukla, J. Kornatowski, D. Demuth, I. Girnus,
H.Pfeifer, L.V.C Rees, S. Schunk, K. Unger and J. K\"arger, 
Science {\bf 272}, 702 (1996).

\bibitem{Wei00}Q-H. Wei, C. Bechinger,  and P. Leiderer, 
Science {\bf 287} 625 (2000).

\bibitem{Lutz04} C. Lutz, M. Kollmann, P. Leiderer, C. Bechinger,
J. Phys. Condens. Matt. {\bf 16}, S4075 (2004).

\bibitem{Coup06}
G. Coupier, M.S. Jean, and C. Guthmann,
cond-mat/0603050 (2006).

\bibitem{Harr65}
T.E. Harris, J. Appl. Prob. {\bf 2}, 323 (1965).

\bibitem{vanB83}
H. van Beijeren, K.W. Kehr, and R. Kutner, 
Phys. Rev. B.  {\bf 28}, 5711 (1983).

\bibitem{Brza05}
A. Brzank and G.M. Sch\"utz,
Appl. Catalysis A {\bf 288}, 194 (2005).

\bibitem{Schu97} 
G.M. Sch\"utz, Int. J. Mod. Phys. B {\bf 11}, 197 (1997).

\bibitem{Basu06}
A. Basu and D. Chowdhury, cond-mat/0608098 (2006).

\bibitem{Nish05}
K. Nishinari, Y. Okada, A. Schadschneider, and D. Chowdhury, 
Phys. Rev. Lett. {\bf 95}, 118101 (2005).

\bibitem{Schu03} 
G.M. Sch\"utz, J. Phys. A {\bf 36}, R339 (2003).

\bibitem{Schu05} 
G.M. Sch\"utz, Diffusion Fundamentals {\bf 2}, 5 (2005).

\bibitem{Fife79} P. Fife,
{\it Mathematical aspects of reacting and diffusing systems}
Lecture Notes in Biomath. {\bf 28} (Springer, Berlin, 1979).

\bibitem{Burg74}
J.M. Burgers, {\it The Non Linear Diffusion Equation} (Reidel, Boston, 1974).

\bibitem{Fish37}
R.A. Fisher,
Ann. Eugenics {\bf 7}, 353  (1937).

\bibitem{Lebo88}
J.L. Lebowitz, E. Presutti, and H. Spohn,
J. Stat. Phys. {\bf 51}, 841 (1988).

\bibitem{Spoh91}
H. Spohn,
{\it Large-Scale Dynamics of Interacting Particles}
(Springer, Berlin, 1991).

\bibitem{Kipn99}
C. Kipnis and C. Landim,
{\it Scaling limits of interacting particle systems}
(Springer, Berlin, 1999).

\bibitem{Ligg99} 
T.M. Liggett, {\em Stochastic Interacting Systems: Voter,
Contact and Exclusion Processes},
(Springer, Berlin, 1999).

\bibitem{Schu00} 
G.M. Sch\"utz, in: {\it Phase Transitions and
Critical Phenomena} Vol. {\bf 19}, C. Domb and J. Lebowitz (eds.),
(Academic, London, 2001).

\bibitem{Ferr91}
       P.A. Ferrari, C. Kipnis, E. Saada, 
Ann. Prob. {\bf 19}, 226 (1991).

\bibitem{Derr97}
      B. Derrida, J.L. Lebowitz, E.R. Speer,
 J. Stat. Phys. {\bf 89}, 135 (1997).

\bibitem{Derr98}
       B. Derrida, S. Goldstein, J.L. Lebowitz, E. R. Speer,
J. Stat. Phys. {\bf 93}, 547 (1998).

\bibitem{Pigo00}
C. Pigorsch,G.M. Sch\"utz, 
J. Phys. A {\bf 33}, 7919 (2000).

\bibitem{Bala01}M. Balazs,
J. Stat. Phys. {\bf 105}, 511 (2001).

\bibitem{Beli02}
      V. Belitsky, G.M. Sch{\"u}tz, 
El. J. Prob. {\bf 7}, 11 (2002).

\bibitem{Kreb03}
K. Krebs, F.H. Jafarpour, and G.M. Sch\"utz, 
New J. Phys. {\bf 5}, 145 (2003).

\bibitem{Rako04} 
A. R\'akos and G.M. Sch\"utz,
J. Stat. Phys. {\bf 117}, 55 (2004).

\bibitem{Bala04}
M. Balazs, 
J. Stat. Phys. {\bf 117}, 77 (2004).

\bibitem{Jafa05}
F. H. Jafarpour
Physica A {\bf 358}, 413 (2005).

\bibitem{Arab06}
M. Arabsalmani, A. Aghamohammadi, 
Phys. Rev. E, {\bf 74}, 011107 (2006).

\bibitem{Doer91}
C.R. Doering, M.A. Burschka, and W. Horsthemke,
J. Stat. Phys. {\bf 65}, 953 (1991).

\bibitem{Hinr96}
H. Hinrichsen, K. Krebs, and I. Peschel,
Z. Phys. B {\bf 100}, 105 (1996).

\bibitem{benA98}
D. ben-Avraham,
Phys. Lett. A {\bf 247}, 53 (1998).

\bibitem{Paes04}
M. Paessens and G.M. Sch\"utz, New. J. Phys. {\bf 6}, 120 (2004).

\bibitem{Glau63} 
R. Glauber, J. Math. Phys. {\bf 4}, 294 (1963).

\bibitem{Krug91} 
J. Krug, Phys. Rev. Lett. {\bf 67}, 1882 (1991).

\bibitem{Schu93}
G. Sch\"utz and E. Domany,
J. Stat. Phys. {\bf 72}, 277 (1993).

\bibitem{Derr93}
B. Derrida, M.R. Evans, V. Hakim, and V. Pasquier,
J. Phys. A {\bf 26}, 1493 (1993).

\bibitem{Kolo98}  
A.B. Kolomeisky, G.M. Sch\"{u}tz, E.B. Kolomeisky, and J.P. Straley,
J. Phys. A {\bf 31}, 6911 (1998).

\bibitem{Popk99} 
V. Popkov and G.M. Sch\"utz,
Europhys. Lett. {\bf 48}, 257 (1999).

\bibitem{Parm03} 
A. Parmeggiani, T. Franosch, and E. Frey,
Phys. Rev. Lett. {\bf 90}, 086601 (2003).

\bibitem{Popk03} 
V. Popkov, A. R\'akos, R. D. Willmann, A. B. Kolomeisky, and G. M. Sch\"utz,
Phys. Rev. E {\bf 67}, 066117 (2003).

\bibitem{Evan03} 
M. R. Evans, R. Juhasz,  L. Santen, 
Phys. Rev. E {\bf 68}, 026117 (2003).

\bibitem{Rako03} 
A. R\'akos, M. Paessens, and G.M. Sch\"utz,
Phys. Rev. Lett. {\bf 91}, 238302 (2003).

\bibitem{Taba06a}
F. Tabatabaei and G.M. Sch\"utz,
cond-mat/0608147 (2006).

\bibitem{Schu96} 
G.M. Sch\"utz, R. Ramaswamy and M. Barma, 
J. Phys. A. {\bf 29}, 837 (1996).

\bibitem{Priv97} V. Privman (ed.), {\em Nonequilibrium Statistical 
Mechanics in One Dimension}, 
(Cambridge, Cambridge University Press, 1997).

\bibitem{Robe84}
M. Robert and B. Widom,
J. Stat. Phys. {\bf 37}, 419 (1984).

\end{thebibliography}

\end{document}